# Laser frequency stabilization by combining modulation transfer and frequency modulation spectroscopy

Fei Zi,[1,2] Xuejian Wu,[1] Weicheng Zhong,[1] Richard H. Parker,[1] Chenghui Yu,[1] Simon Budker,[1] Xuanhui Lu,[2] Holger Müller[1, *]

1 Department of Physics, 366 Le Conte Hall MS 7300, University of California, Berkeley, California 94720, USA
2 Institute of Optics, Department of Physics, Zhejiang University, Hangzhou, China
*Corresponding author: hm@berkeley.edu

We present a hybrid laser frequency stabilization method combining modulation transfer spectroscopy (MTS) and frequency modulation spectroscopy (FMS) for the cesium $D_2$ transition. In a typical pump-probe setup, the error signal is a combination of the DC-coupled MTS error signal and the AC-coupled FMS error signal. This combines the long-term stability of the former with the high signal-to-noise ratio of the latter. In addition, we enhance the long-term frequency stability with laser intensity stabilization. By measuring the frequency difference between two independent hybrid spectroscopies, we investigate the short-term and long-term stability. We find a long-term stability of 7.8 kHz characterized by a standard deviation of the beating frequency drift over the course of 10 hours, and a short-term stability of 1.9 kHz characterized by an Allan deviation of that at 2 seconds of integration time.

## 1. INTRODUCTION

Laser frequency stabilization is critical in atomic physics [1-3]. Doppler-free techniques, such as saturated absorption spectroscopy [4-7], are commonly used to reference laser frequencies to atomic transitions. If the laser frequency is close to a transition, two counter-propagating beams interact with atoms of the same velocity class, and Doppler broadening due to the thermal velocity distribution of the atoms is suppressed. More specifically, spectroscopy techniques can be classified as either unmodulated or modulated. Typical unmodulated methods include dichroic atomic vapor laser locking [8] and polarization spectroscopy [9-12]. Without modulation, the optical setup can be compact, but the error signals often suffer from a stronger DC offset drift, which causes frequency instability. In contrast, modulated spectroscopy, such as frequency modulation spectroscopy (FMS) [13-16] and modulation transfer spectroscopy (MTS) [17-20], detect error signals at higher frequencies above the technical noise floor of the laser, which makes the error signal less sensitive to the DC offset drift.

In FMS, the signal on the modulated beam is demodulated directly, which leads to a good signal-to-noise ratio, but also to a DC drift that depends on, for example, residual amplitude modulation, which cannot be completely removed. In MTS, the error signal comes from demodulating a signal generated by four-wave mixing in the initially unmodulated beam. This suppresses background signals and typically leads to high long-term stability. By taking advantage of the better short-term performance of the FMS and the better long-term performance of the MTS, it is possible to further improve the stability of the laser frequency. The long-term frequency stability of the spectroscopy is important for precision measurements. For example, fine-structure measurements using atom interferometry require the laser frequency drift to be less than 100 kHz in order to achieve an uncertainty of 0.25 ppb [21]. In addition, precision measurements, such as atom interferometers, require hours or even days of data acquisition to increase the accuracy, so reliable spectroscopy with stable long-term performance is essential as a flywheel oscillator, even when an optical frequency comb is available for calibration.

In this paper, we demonstrate hybrid spectroscopy by combining the error signals from FMS and MTS in a typical spectroscopy setup. Both the pump beam and the probe beam are detected, and the hybrid error signal is generated by summing the AC-coupled FMS error signal and the DC-coupled MTS error signal. Hybrid spectroscopy combines the improved signal-to-noise ratio of FMS with the background-free advantage of MTS. The short-term and long-term stability of the hybrid spectroscopy is evaluated by measuring the frequency variation of the beat note between these two independent hybrid spectroscopies.

## 2. EXPERIMENTAL SETUP

Figure 1 shows a schematic of our hybrid spectroscopy. A 100-MHz acousto-optic modulator (AO) acts as a beam splitter to generate the pump and probe beams. Giving the pump beam and the probe beam different frequencies with the AO, parasitic interferences caused by the reflections from either the cesium vapor cell or the photo-detectors are eliminated. The pump beam is phase modulated at 9.2 MHz by an electro-optic modulator (Thorlabs, EO-PM-NR-C1), whose modulation phase index is ~0.8. A 75-mm-long cesium vapor cell (Triad Technology Inc.), whose vapor pressure is ~1.5×10$^{-6}$ torr at 25 °C, is placed inside a magnetic shield. The laser power through the vapor cell is approximately 1 mW. A telescope before the vapor cell increases the beam size to ~3.5 mm (1/$e^2$ intensity radius). The signal of the MTS is detected by photo-detector D1 while the FMS signal is detected by photo-detector D2. The signals are demodulated by mixing with the oscillator. A phase shifter is inserted before combining with the error signal of the MTS, in order to ensure that the two error signals have the same phase. The FMS signal is AC-coupled through a 2-μF capacitor while the MTS signal is DC-coupled through a 10 kΩ resistor. The error signal of the hybrid spectroscopy is delivered to a servo, and feedback is given to both the current and the piezo of the external cavity diode laser (New Focus, TLB7115).

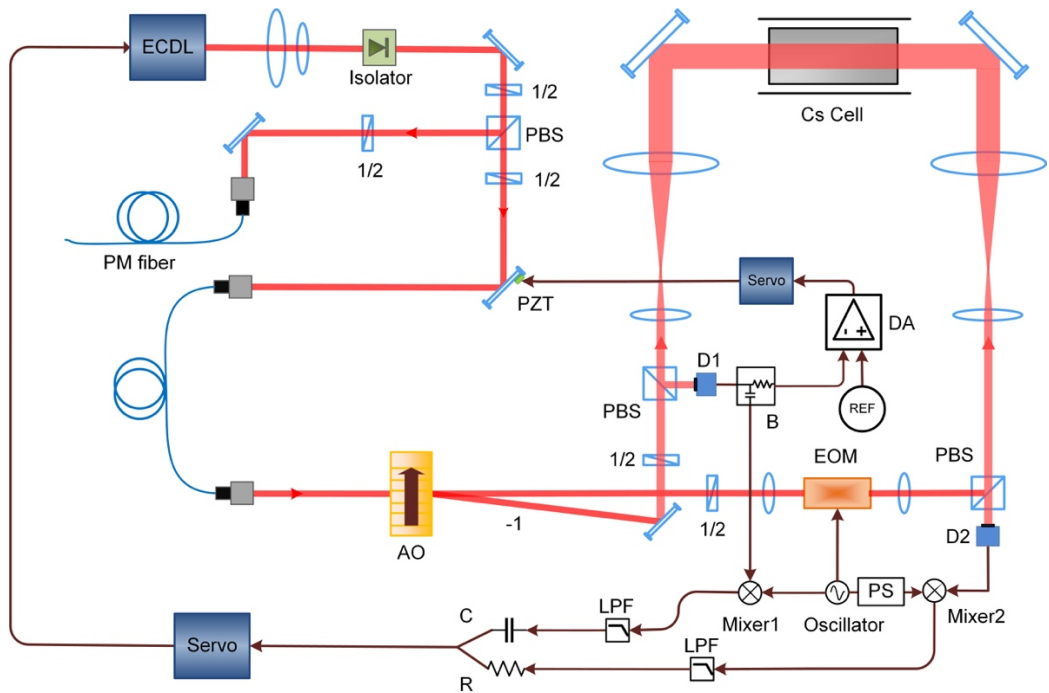

Fig. 1. Schematic of the hybrid spectroscopy. ECDL, external cavity diode laser; PBS, polarization beam splitter; PM fiber, polarization maintaining fiber; AO, acousto-optic modulator; EOM, electro-optic modulator; D1 and D2, photo-detector; 1/2, half wave plate; LPF, low pass filter; PS, phase shifter; B, bias-tee; DA, differential amplifier; REF, reference voltage; PZT, piezoelectric actuator; R, 10-kΩ resistor; C, 2-μF capacitor.

In order to stabilize the laser intensity, the mirror mount for the laser coupling to the spectroscopy is installed with a piezoelectric actuator (Thorlabs, AE0203D08F), which is fast enough to compensate for the slow intensity drift. The DC output from D2 serves as a laser-intensity indicator. Subtracting a constant DC voltage from the voltage of D2 produces the resulting error signal, which is sent to another servo with a high-voltage amplifier.

In the experiment performed here, the laser frequency is set close to the $F=3 \rightarrow F'=2$ transition of cesium $D_2$ transition. The error signals of the FMS and MTS before combining are shown in figure 2. Before combining, the amplitudes of the FMS and MTS error signals are set to approximately the same value. Since the error signal of the hybrid spectroscopy consists of the DC-coupled MTS error signal and the AC-coupled FMS error signal, the characteristics of background-free from MTS and high signal-to-noise ratio from FMS are both obtained.

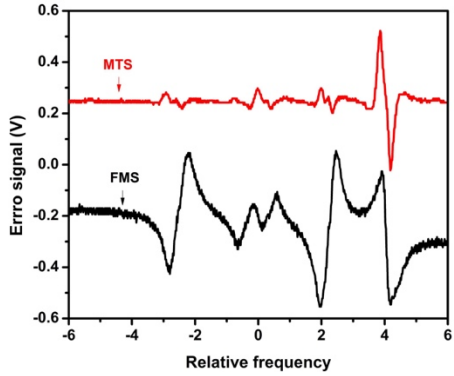

Fig. 2. Error signals of FMS and MTS before combination. The MTS and FMS are offset by +0.25 V and -0.25 V, respectively.

## 3. RESULTS

In order to investigate the frequency stability, we measure the frequency variation of the beat note between two independent hybrid spectroscopies for 10 hours. Both hybrid spectroscopies are locked to the $F=3 \rightarrow F'=2$ transition, and the beat frequency is set to approximately 300 MHz by using additional AOs. To ensure the long-term frequency stability, the laser intensity is stabilized before the spectroscopy. We find that the frequency of the beat signal changed by ~160 kHz if the laser power before the spectroscopy changes by ~0.2 mW, which corresponds to ~120 mV variation of the D2 output. With intensity stabilization, the variation of the DC voltage from D2 is controlled to within 5 mV, implying that such frequency change has been reduced to within 8 kHz.

A frequency counter, synchronized to a 10 MHz reference, is used to measure the beat frequency with a gate time of 1 s. Since the two hybrid spectroscopies run independently under the similar environment, we can assume that the two spectroscopies contribute equally to the frequency variation. Therefore, the frequency stability of either one should be divided by $\sqrt{2}$ from the frequency stability of the beat note. Figure 3a shows the variation of the beat frequency, which has a standard deviation of 11 kHz, and a peak-to-peak drift of 82 kHz. Thus, the standard deviation of either hybrid spectroscopy is 7.8 kHz over 10 hours. Figure 3b shows the Allan deviation of the beat frequency, and indicates that the short-term stability is 2.7 kHz with an averaging time of 2 seconds. For either hybrid spectroscopy, the Allan deviation at 2-second averaging time is 1.9 kHz. For comparison, the beat notes are also measured when both spectroscopies work as MTS or FMS individually, and the Allan deviations are shown in Figure 3b as well. The hybrid spectroscopy shows better short-term and better long-term frequency stability than either MTS or FMS respectively.

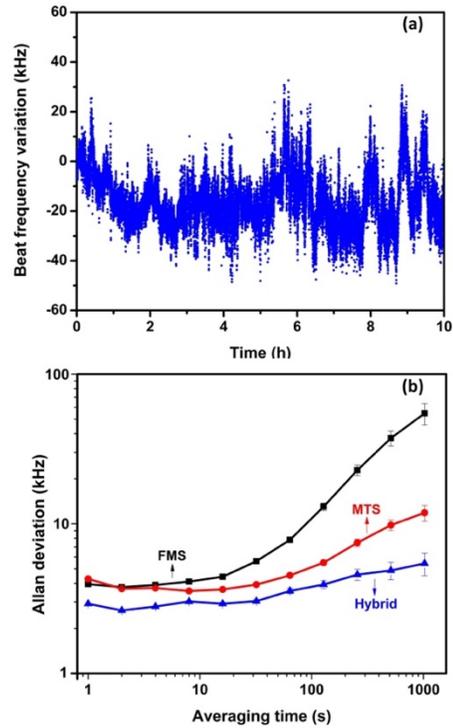

Fig. 3. (a) Beat frequency variation of two independent hybrid spectroscopies. The standard deviation and the peak-to-peak drifts are 11 kHz and 82 kHz, respectively. The gate time of the frequency measurement is 1 s. (b) Allan variance of the beat frequency when both spectroscopies work as hybrid, MTS, or FMS individually. For the hybrid spectroscopy, the best Allan deviation is 2.7 kHz at an averaging time of 2 seconds.

## 4. CONCLUSION

In conclusion, the hybrid spectroscopy combines the superior short-term performance of FMS and the superior long-term performance of MTS. Frequency stability is improved in a relatively easy and compact manner. We investigate the frequency stability by measuring the difference frequency between two independent hybrid spectroscopies over 10 hours. The long-term stability is characterized by a standard deviation of 7.8 kHz, and the short-term stability is characterized by an Allan deviation of 1.9 kHz at a 2-second integration time. The long-term stability of this hybrid spectroscopy is 5 times as good as that of the reported unmodulated polarization spectroscopy [12], and twice as good as the typical modulation transfer spectroscopy [20].

**Funding Information.** National Science Foundation under CAREER Grant (PHY-1056620); The David and Lucile Packard Foundation; The Israeli Ministry of Defense; National Aeronautics and Space Administration Grants (NNH13ZTT002N, NNH10ZDA001N-PIDDP, and NNH11ZTT001); National Science Foundation of China (NSFC) (11474254); The 973 Program (2012CB921602); Fundamental Research Funds for the Central Universities of China (2016XZZX004-01)

**Acknowledgment**. F. Z. wants to thank the fellowships from Zhejiang University and Tang Lixin for supporting her visiting researches at UC Berkeley.